\documentclass[12pt,twoside]{article}
\usepackage[english]{babel}
\usepackage{graphicx,rotating,epsfig}
\usepackage{a4p}
\usepackage{xspace}

\def\ra{\rightarrow}
\def\GeV  {\ensuremath{\mathrm{ Ge\kern -0.1em V } }}
\def\GeVc2{\ensuremath{\mathrm{ Ge\kern -0.1em V }\kern -0.2em /c^2 }}
\def\MeVc2{\ensuremath{\mathrm{ Me\kern -0.1em V }\kern -0.2em /c^2 }}
\newcommand{\MT}{\ensuremath{M_{\mathrm{t}}}}
\newcommand{\MTll}{\ensuremath{\MT^{\mathrm{di-l}}}}
\newcommand{\MTlj}{\ensuremath{\MT^{\mathrm{l+j}}}}
\newcommand{\MTjj}{\ensuremath{\MT^{\mathrm{all-j}}}}

\newcommand{\pb}{\ensuremath{\mathrm{pb}^{-1}}}
\newcommand{\fb}{\ensuremath{\mathrm{fb}^{-1}}}
\newcommand{\ttbar}{\ensuremath{t\overline{t}}}
\newcommand{\ljt}{\ensuremath{\ell\nu q q^{\prime} b \overline{b}}}
\newcommand{\had}{\ensuremath{q q^{\prime} b q q^{\prime} \overline{b}}}
\newcommand{\dil}{\ensuremath{\ell^{+}\nu b\ell^{-}\overline{\nu}\overline{b}}}
\newcommand{\ttljt}{\ensuremath{\ttbar\ra\ljt}}
\newcommand{\ttdil}{\ensuremath{\ttbar\ra\dil}}
\newcommand{\tthad}{\ensuremath{\ttbar\ra\had}}
\newcommand{\RunI}{\hbox{Run~I}}
\newcommand{\RunII}{\hbox{Run~II}}
\newcommand{\measStatSyst}[3]{\ensuremath{#1 \pm #2~(\textrm{stat.}) \pm #3~(\textrm{syst.})}\xspace}
\newcommand{\gevcc}[1]  {\ensuremath{#1~\mathrm{GeV}/c^{2}}}

\begin{document}

\begin{center}
  {\LARGE FERMI NATIONAL ACCELERATOR LABORATORY}
\end{center}

\begin{flushright}
       FERMILAB-TM-2413-E \\  
       TEVEWWG/top 2008/02 \\
       CDF Note 9449 \\
       D\O\ Note 5751 \\
       \vspace*{0.05in}
       July 2008 \\
\end{flushright}

\vskip 1cm

\begin{center}
  {\LARGE\bf 
    Combination of CDF and D\O\ Results \\
    on the Mass of the Top Quark\\
  }
  \vfill
  {\Large
    The Tevatron Electroweak Working Group\footnote{The Tevatron Electroweak 
    Working Group can be contacted at tev-ewwg@fnal.gov.\\  
    \hspace*{0.20in} More information can
    be found at {\tt http://tevewwg.fnal.gov}.} \\
    for the CDF and D\O\ Collaborations\\
  }
\end{center}
\vfill
\begin{abstract}
\noindent
  We summarize the top-quark mass measurements from the CDF and
  D\O\ experiments at Fermilab.  We combine published
  \RunI\ (1992-1996) measurements with the most recent preliminary
  \RunII\ (2001-present) measurements using up to $2.8~\fb$ of data.
  Taking correlated uncertainties properly into account the resulting
  preliminary world average mass of the top quark is
  $\MT=\gevcc{\measStatSyst{172.4}{0.7}{1.0}}$,
  assuming Gaussian systematic uncertainties. Adding in quadrature
  yields a total uncertainty of $\gevcc{1.2}$, corresponding to a
  relative precision of 0.7\% on the top-quark mass.
\end{abstract}

\vfill



\section{Introduction}
\label{sec:intro}

The experiments CDF and D\O, taking data at the Tevatron
proton-antiproton collider located at the Fermi National Accelerator
Laboratory, have made several direct experimental measurements of the
top-quark pole mass, \MT.  The pioneering measurements were based on about
$100~\pb$ of \RunI\ (1992-1996) data~\cite{Mtop1-CDF-di-l-PRLa, 
  Mtop1-CDF-di-l-PRLb,
  Mtop1-CDF-di-l-PRLb-E, Mtop1-D0-di-l-PRL, Mtop1-D0-di-l-PRD,
  Mtop1-CDF-l+j-PRL, Mtop1-CDF-l+j-PRD, Mtop1-D0-l+j-old-PRL,
  Mtop1-D0-l+j-old-PRD, Mtop1-D0-l+j-new1, Mtop1-CDF-all-j-PRL,
  Mtop1-D0-all-j-PRL} 
and include results from the \tthad\ (all-j), the \ttljt\ (l+j), and the 
\ttdil\ (di-l) decay channels\footnote{Here $\ell=e$ or $\mu$.  Decay 
channels with explicit tau lepton identification are presently under 
study and are not yet used for measurements of the top-quark mass.}.   
The \RunII\ measurements summarized here are the most recent results in the 
l+j, di-l,  and all-j channels using $1.9-2.8~\fb$ of data and improved 
analysis techniques~\cite{
Mtop2-CDF-di-l-new,
Mtop2-CDF-l+j-new,
Mtop2-CDF-all-j-new, 
Mtop2-CDF-trk-new,
Mtop2-D0-l+ja-final,
Mtop2-D0-l+j-new,
Mtop2-D0-di-l-jul08-1,
Mtop2-D0-di-l-jul08-2}.  
\vspace*{0.10in}

This note reports the world average top-quark mass obtained by
combining five published
\RunI\ measurements~\cite{Mtop1-CDF-di-l-PRLb, Mtop1-CDF-di-l-PRLb-E,
  Mtop1-D0-di-l-PRD, Mtop1-CDF-l+j-PRD, Mtop1-D0-l+j-new1,
  Mtop1-CDF-all-j-PRL} with four preliminary \RunII\ CDF
results~\cite{Mtop2-CDF-di-l-new, Mtop2-CDF-l+j-new, Mtop2-CDF-all-j-new,
Mtop2-CDF-trk-new} and three preliminary
\RunII\ D\O\ results~\cite{Mtop2-D0-l+ja-final, Mtop2-D0-l+j-new,
Mtop2-D0-di-l-jul08-1, Mtop2-D0-di-l-jul08-2}.
The combination takes into account the
statistical and systematic uncertainties and their correlations using
the method of references~\cite{Lyons:1988, Valassi:2003} and
supersedes previous
combinations~\cite{Mtop1-tevewwg04,Mtop-tevewwgSum05,
  Mtop-tevewwgWin06,Mtop-tevewwgSum06, Mtop-tevewwgWin07, Mtop-tevewwgWin08}.
\vspace*{0.10in}

The input measurements and error categories used in the combination are 
detailed in Sections~\ref{sec:inputs} and~\ref{sec:errors}, respectively. 
The correlations used in the combination are discussed in 
Section~\ref{sec:corltns} and the resulting world average top-quark mass 
is given in Section~\ref{sec:results}.  A summary and outlook are presented
in Section~\ref{sec:summary}.
 
\section{Input Measurements}
\label{sec:inputs}

For this combination twelve measurements of \MT\ are used: five
published \RunI\ results, and seven preliminary
\RunII\ results, all reported in Table~\ref{tab:inputs}.  In general,
the \RunI\ measurements all have relatively large statistical
uncertainties and their systematic uncertainty is dominated by the
total jet energy scale (JES) uncertainty.  In \RunII\ both CDF and
D\O\ take advantage of the larger \ttbar\ samples available and employ
new analysis techniques to reduce both these uncertainties.  In
particular the JES is constrained using an in-situ calibration based
on the invariant mass of $W\ra qq^{\prime}$ decays in the l+j and
all-j channels.  The \RunII\ D\O\ analysis in the l+j channel
constrains the response of light-quark jets using the in-situ $W\ra
qq^{\prime}$ decays.  Residual JES uncertainties associated with
$\eta$ and $p_{T}$ dependencies as well as uncertainties specific to
the response of $b$-jets are treated separately.  Similarly, the
\RunII\ CDF analyses in the l+j and all-j channels also constrain the
JES using the in-situ $W\ra qq^{\prime}$ decays.  Small residual JES
uncertainties arising from $\eta$ and $p_{T}$ dependencies and the
modeling of $b$-jets are included in separate error categories.  The
\RunII\ CDF and D\O\ di-l measurements use a JES determined from external
calibration samples.  Some parts of the associated uncertainty are
correlated with the \RunI\ JES uncertainty as noted below.
\vspace*{0.10in}

\begin{table}[t]
\begin{center}
\renewcommand{\arraystretch}{1.30}
{\small
\begin{tabular}{|l||rrr|rr||rrrr|rrr|}
\hline       
       & \multicolumn{5}{|c||}{{\RunI} published} & \multicolumn{7}{|c|}{{\RunII} preliminary} \\ \cline{2-13}
       & \multicolumn{3}{|c|}{ CDF } & \multicolumn{2}{|c||}{ D\O\ }
       & \multicolumn{4}{|c|}{ CDF } & \multicolumn{3}{|c|}{ D\O\ } \\
       & all-j & l+j   & di-l  & l+j   & di-l  & l+j   & di-l  & all-j & trk & l+j/a & l+j/b & di-l \\
\hline
$\int \mathcal{L}\;dt$ & 0.1 & 0.1 & 0.1 & 0.1 & 0.1 & 2.7 & 1.9 & 2.1 & 1.9 & 1.0 & 1.2 & 2.8 \\
\hline
\hline                         
Result & 186.0 & 176.1 & 167.4 & 180.1 & 168.4 & 172.2 & 171.2 & 176.9 & 175.3 & 171.5 & 173.0 & 174.4 \\
\hline                         
\hline                         
iJES   &   0.0 &   0.0 &   0.0 &   0.0 &   0.0 &   0.9 &   0.0 &   1.9 &   0.0 &   0.7 &   1.4 &   0.0 \\
aJES   &   0.0 &   0.0 &   0.0 &   0.0 &   0.0 &   0.0 &   0.0 &   0.0 &   0.0 &   0.8 &   0.8 &   1.2 \\
bJES   &   0.6 &   0.6 &   0.8 &   0.7 &   0.7 &   0.4 &   0.4 &   0.6 &   0.0 &   0.0 &   0.1 &   0.3 \\
cJES   &   3.0 &   2.7 &   2.6 &   2.0 &   2.0 &   0.3 &   1.7 &   0.6 &   0.6 &   0.0 &   0.0 &   0.0 \\
dJES   &   0.3 &   0.7 &   0.6 &   0.0 &   0.0 &   0.0 &   0.1 &   0.1 &   0.0 &   0.8 &   0.0 &   1.6 \\
rJES   &   4.0 &   3.4 &   2.7 &   2.5 &   1.1 &   0.4 &   1.9 &   0.6 &   0.1 &   0.0 &   0.0 &   0.0 \\
lepPt  &   0.0 &   0.0 &   0.0 &   0.0 &   0.0 &   0.2 &   0.1 &   0.0 &   1.1 &   0.0 &   0.0 &   0.0 \\
Signal &   1.8 &   2.6 &   2.8 &   1.1 &   1.8 &   0.3 &   0.8 &   0.5 &   1.6 &   0.5 &   0.5 &   0.5 \\
BG     &   1.7 &   1.3 &   0.3 &   1.0 &   1.1 &   0.4 &   0.4 &   1.0 &   1.6 &   0.4 &   0.4 &   0.6 \\
Fit    &   0.6 &   0.0 &   0.7 &   0.6 &   1.1 &   0.2 &   0.6 &   0.5 &   1.4 &   0.3 &   0.2 &   0.3 \\
MC     &   0.8 &   0.1 &   0.6 &   0.0 &   0.0 &   0.5 &   0.9 &   0.5 &   0.6 &   0.0 &   0.0 &   0.0 \\
UN/MI  &   0.0 &   0.0 &   0.0 &   1.3 &   1.3 &   0.0 &   0.0 &   0.0 &   0.0 &   0.0 &   0.0 &   0.0 \\
\hline                         
Syst.  &   5.7 &   5.3 &   4.9 &   3.9 &   3.6 &   1.3 &   2.9 &   2.6 &   3.0 &   1.5 &   1.7 &   2.1 \\
Stat.  &  10.0 &   5.1 &  10.3 &   3.6 &  12.3 &   1.0 &   2.7 &   3.3 &   6.2 &   1.5 &   1.3 &   3.2 \\
\hline                         
\hline                         
Total  &  11.5 &   7.3 &  11.4 &   5.3 &  12.8 &   1.7 &   4.0 &   4.2 &   6.9 &   2.1 &   2.2 &   3.9 \\ 
\hline
\end{tabular}
}
\end{center}
\caption[Input measurements]{Summary of the measurements used to determine the
  world average $\MT$.  Integrated luminosity ($\int \mathcal{L}\;dt$) is in
  \fb, and all other numbers are in $\GeVc2$.  The error categories and 
  their correlations are described in the text.  The total systematic uncertainty 
  and the total uncertainty are obtained by adding the relevant contributions 
  in quadrature.}
\label{tab:inputs}
\end{table}


The D\O\ Run~IIa l+j analysis is using the JES determined from the
external calibration derived using $\gamma$+jets events as an
additional Gaussian constraint to the in-situ calibration. Therefore
the total resulting JES uncertainty has been split into the part
coming solely from the in-situ calibration and the part coming from
the external calibration. To do that, the measurement without external
JES constraint has been combined iteratively with a pseudo-measurement
using the BLUE method that would use only the external calibration so
that the combination gives the actual total JES uncertainty. The
splitting obtained in this way is used to assess the iJES and dJES
uncertainties~\cite{Mtop2-D0-comb}.

A new analysis technique from CDF is included (trk).
This measurement uses both the mean
decay-length from B-tagged jets and the mean lepton transverse momentum
to determine the top-quark mass in l+j candidate events.
While the statistical sensitivity is not as good as the more
traditional methods, this technique has the advantage that since it
uses primarily tracking information, it is almost entirely independent of
JES uncertainties.  As the statitistics of this sample continue to
grow, this method could offer a nice cross-check of the top-quark mass
that's largely independent of the dominant JES systematic uncertainty
which plagues the other measurements.  The statistical correlation
between an earlier version of the trk analysis and a
traditional \RunII\ CDF l+j measurement was
studied using Monte Carlo signal-plus-background pseudo-experiments
which correctly account for the sample overlap and was found to be
consistent with zero (to within $<1\%$) independent of the assumed
top-quark mass.
\vspace*{0.10in}

The two D\O\ \RunII\ lepton+jets results~\cite{Mtop2-D0-l+ja-final,
Mtop2-D0-l+j-new} are derived from Run~IIa and Run~IIb datasets,
respectively, and are labelled as such.  The D\O\ \RunII\ dilepton
result is itself a combination of two results
using different techniques but partially overlapping dilepton data
sets~\cite{Mtop2-D0-di-l-jul08-1,Mtop2-D0-di-l-jul08-2}.
\vspace*{0.10in}

Table~\ref{tab:inputs} also lists the uncertainties of the results,
sub-divided into the categories described in the next Section.  The
correlations between the inputs are described in
Section~\ref{sec:corltns}.


\section{Error Categories}
\label{sec:errors}

We employ the same error categories as used for the previous world
average~\cite{Mtop-tevewwgWin08}, plus one new category (lepPt).  They
include a detailed breakdown of the various sources of uncertainty and
aim to lump together sources of systematic uncertainty that share the
same or similar origin.  For example, the ``Signal'' category
discussed below includes the uncertainties from ISR, FSR, and
PDF---all of which affect the modeling of the \ttbar\ signal.  Some
systematic uncertainties have been broken down into multiple
categories in order to accommodate specific types of correlations.
For example, the jet energy scale (JES) uncertainty is sub-divided
into several components in order to more accurately accommodate our
best estimate of the relevant correlations.  Each error category is
discussed below.
\vspace*{0.10in}

\begin{description}
  \item[Statistical:] The statistical uncertainty associated with the
    \MT\ determination.
 \item[iJES:] That part of the JES uncertainty which originates from
   in-situ calibration procedures and is uncorrelated among the
   measurements.  In the combination reported here it corresponds to
   the statistical uncertainty associated with the JES determination
   using the $W\ra qq^{\prime}$ invariant mass in the CDF \RunII\
   l+j and all-h measurements and D\O\ Run~IIa and Run~IIb l+j
   measurements. Residual JES uncertainties, which arise
   from effects
   not considered in the in-situ calibration, are included in other
   categories.
  \item[aJES:] That part of the JES uncertainty which originates from
    differences in detector $e/h$ response between $b$-jets and light-quark
    jets.  It is specific to the D\O\ \RunII\ measurements and is
    taken to be uncorrelated with the D\O\ \RunI\ and CDF measurements.
  \item[bJES:] That part of the JES uncertainty which originates from
    uncertainties specific to the modeling of $b$-jets and which is correlated
    across all measurements.  For both CDF and D\O\ this includes uncertainties 
    arising from 
    variations in the semi-leptonic branching fraction, $b$-fragmentation 
    modeling, and differences in the color flow between $b$-jets and light-quark
    jets.  These were determined from \RunII\ studies but back-propagated
    to the \RunI\ measurements, whose rJES uncertainties (see below) were 
    then corrected in order to keep the total JES uncertainty constant.
  \item[cJES:] That part of the JES uncertainty which originates from
    modeling uncertainties correlated across all measurements.  Specifically
    it includes the modeling uncertainties associated with light-quark 
    fragmentation and out-of-cone corrections.
  \item[dJES:] That part of the JES uncertainty which originates from
   limitations in the calibration data samples used and which is
   correlated between measurements within the same data-taking
   period, such as \RunI\ or \RunII, but not between
   experiments.  For CDF this corresponds to uncertainties associated
   with the $\eta$-dependent JES corrections which are estimated
   using di-jet data events. For D\O\ this includes uncertainties in the
   calorimeter response to light-quark jets, and $\eta$- and
   $p_{T}$-dependent uncertainties
   constrained using \RunII\ $\gamma+$jet data samples.
  \item[rJES:] The remaining part of the JES uncertainty which is 
    correlated between all measurements of the same experiment 
    independent of data-taking period, but is uncorrelated between
    experiments.  For CDF, this is dominated by uncertainties in the
    calorimeter response to light-quark jets, and also includes small 
    uncertainties associated with the multiple interaction and underlying 
    event corrections.
  \item[lepPt:] The systematic uncertainty arising from uncertainties
    in the scale of lepton transverse momentum measurements.  This is an
    important uncertainty for CDF's track-based measurement.  It was not
    considered as a source of systematic uncertainty in the \RunI\
    measurements or in measurements at D\O.
  \item[Signal:] The systematic uncertainty arising from uncertainties
    in the modeling of the \ttbar\ signal which is correlated across all
    measurements.  This includes uncertainties from variations in the ISR,
    FSR, and PDF descriptions used to generate the \ttbar\ Monte Carlo samples
    that calibrate each method.  It also includes small uncertainties 
    associated with biases associated with the identification of $b$-jets.
  \item[Background:]  The systematic uncertainty arising from uncertainties
    in modeling the dominant background sources and correlated across
    all measurements in the same channel.  These
    include uncertainties on the background composition and shape.  In
    particular uncertainties associated with the modeling of the QCD
    multi-jet background (all-j and l+j), uncertainties associated with the
    modeling of the Drell-Yan background (di-l), and uncertainties associated 
    with variations of the fragmentation scale used to model W+jets 
    background (all channels) are included.
  \item[Fit:] The systematic uncertainty arising from any source specific
    to a particular fit method, including the finite Monte Carlo statistics 
    available to calibrate each method.
  \item[Monte Carlo:] The systematic uncertainty associated with variations
    of the physics model used to calibrate the fit methods and correlated
    across all measurements.  For CDF it includes variations observed when 
    substituting PYTHIA~\cite{PYTHIA4,PYTHIA5,PYTHIA6} (\RunI\ and \RunII) 
    or ISAJET~\cite{ISAJET} (\RunI) for HERWIG~\cite{HERWIG5,HERWIG6} when 
    modeling the \ttbar\ signal.  Similar
    variations are included for the D\O\ \RunI\ measurements.  The D\O\ 
    \RunII\ measurements use ALPGEN~\cite{ALPGEN} to model the \ttbar\ signal and the
    variations considered are included in the Signal category above.
  \item[UN/MI:] This is specific to D\O\ and includes the uncertainty
    arising from uranium noise in the D\O\ calorimeter and from the
    multiple interaction corrections to the JES.  For D\O\ \RunI\ these
    uncertainties were sizable, while for \RunII, owing to the shorter
    integration time and in-situ JES determination, these uncertainties
    are negligible.
\end{description}
These categories represent the current preliminary understanding of the
various sources of uncertainty and their correlations.  We expect these to 
evolve as we continue to probe each method's sensitivity to the various 
systematic sources with ever improving precision.  Variations in the assignment
of uncertainties to the error categories, in the back-propagation of the bJES
uncertainties to \RunI\ measurements, in the approximations made to
symmetrize the uncertainties used in the combination, and in the assumed 
magnitude of the correlations all negligibly effect ($\ll 0.1\GeVc2$) the 
combined \MT\ and total uncertainty.

\section{Correlations}
\label{sec:corltns}

The following correlations are used when making the combination:
\begin{itemize}
  \item The uncertainties in the Statistical, Fit, and iJES
    categories are taken to be uncorrelated among the measurements.
  \item The uncertainties in the aJES and dJES categories are taken
    to be 100\% correlated among all \RunI\ and all \RunII\ measurements 
    on the same experiment, but uncorrelated between \RunI\ and \RunII\
    and uncorrelated between the experiments.
  \item The uncertainties in the rJES and UN/MI categories are taken
    to be 100\% correlated among all measurements on the same experiment.
  \item The uncertainties in the Background category are taken to be
    100\% correlated among all measurements in the same channel.
  \item The uncertainties in the bJES, cJES, Signal, and Generator
    categories are taken to be 100\% correlated among all measurements.
\end{itemize}
Using the inputs from Table~\ref{tab:inputs} and the correlations specified
here, the resulting matrix of total correlation co-efficients is given in
Table~\ref{tab:coeff}.

\begin{table}[t]
\begin{center}
\renewcommand{\arraystretch}{1.30}
\begin{tabular}{|ll||rrr|rr||rrrr|rrr|}
\hline       
   &   & \multicolumn{5}{|c||}{{\RunI} published} & \multicolumn{7}{|c|}{{\RunII} preliminary} \\ \cline{3-14}
   &   & \multicolumn{3}{|c|}{ CDF } & \multicolumn{2}{|c||}{ D\O\ }
       & \multicolumn{4}{|c|}{ CDF } & \multicolumn{3}{|c|}{ D\O\ } \\
   &            & l+j & di-l & all-j &   l+j &  di-l & l+j   & di-l  & all-j & trk & l+j/a & l+j/b & di-l \\
\hline
\hline
CDF-I & l+j     & 1.00&      &       &       &       &       &       &      &      &      &      & \\
CDF-I & di-l    & 0.29&  1.00&       &       &       &       &       &      &      &      &      & \\
CDF-I & all-j   & 0.32&  0.19&   1.00&       &       &       &       &      &      &      &      & \\
\hline
D\O-I & l+j     & 0.26&  0.15&   0.14&   1.00&       &       &       &      &      &      &      & \\
D\O-I & di-l    & 0.11&  0.08&   0.07&   0.16&   1.00&       &       &      &      &      &      & \\
\hline
\hline
CDF-II & l+j    & 0.32&  0.18&   0.20&   0.19&   0.07&   1.00&       &      &      &      &      & \\
CDF-II & di-l   & 0.45&  0.28&   0.33&   0.22&   0.11&   0.34&   1.00&      &      &      &      & \\
CDF-II & all-j  & 0.17&  0.11&   0.16&   0.10&   0.05&   0.15&   0.19&  1.00&      &      &      & \\
CDF-II & trk    & 0.16&  0.08&   0.07&   0.13&   0.05&   0.17&   0.12&  0.05&  1.00&      &      & \\
\hline
D\O-II & l+j/a  & 0.11&  0.05&   0.03&   0.08&   0.03&   0.09&   0.04&  0.03&  0.09&  1.00&      & \\
D\O-II & l+j/b  & 0.12&  0.06&   0.04&   0.09&   0.03&   0.10&   0.05&  0.03&  0.09&  0.24& 1.00 & \\
D\O-II & di-l   & 0.05&  0.04&   0.02&   0.04&   0.04&   0.04&   0.05&  0.03&  0.03&  0.31& 0.16 & 1.00\\
\hline
\end{tabular}
\end{center}
\caption[Global correlations between input measurements]{The resulting
  matrix of total correlation coefficients used to determined the
  world average top quark mass.}
\label{tab:coeff}
\end{table}

The measurements are combined using a program implementing a numerical
$\chi^2$ minimization as well as the analytic BLUE
method~\cite{Lyons:1988, Valassi:2003}. The two methods used are
mathematically equivalent, and are also equivalent to the method used
in an older combination~\cite{TM-2084}, and give identical results for
the combination. In addition, the BLUE method yields the decomposition
of the error on the average in terms of the error categories specified
for the input measurements~\cite{Valassi:2003}.

\section{Results}
\label{sec:results}

\begin{figure}[p]
\begin{center}
\includegraphics[width=0.8\textwidth]{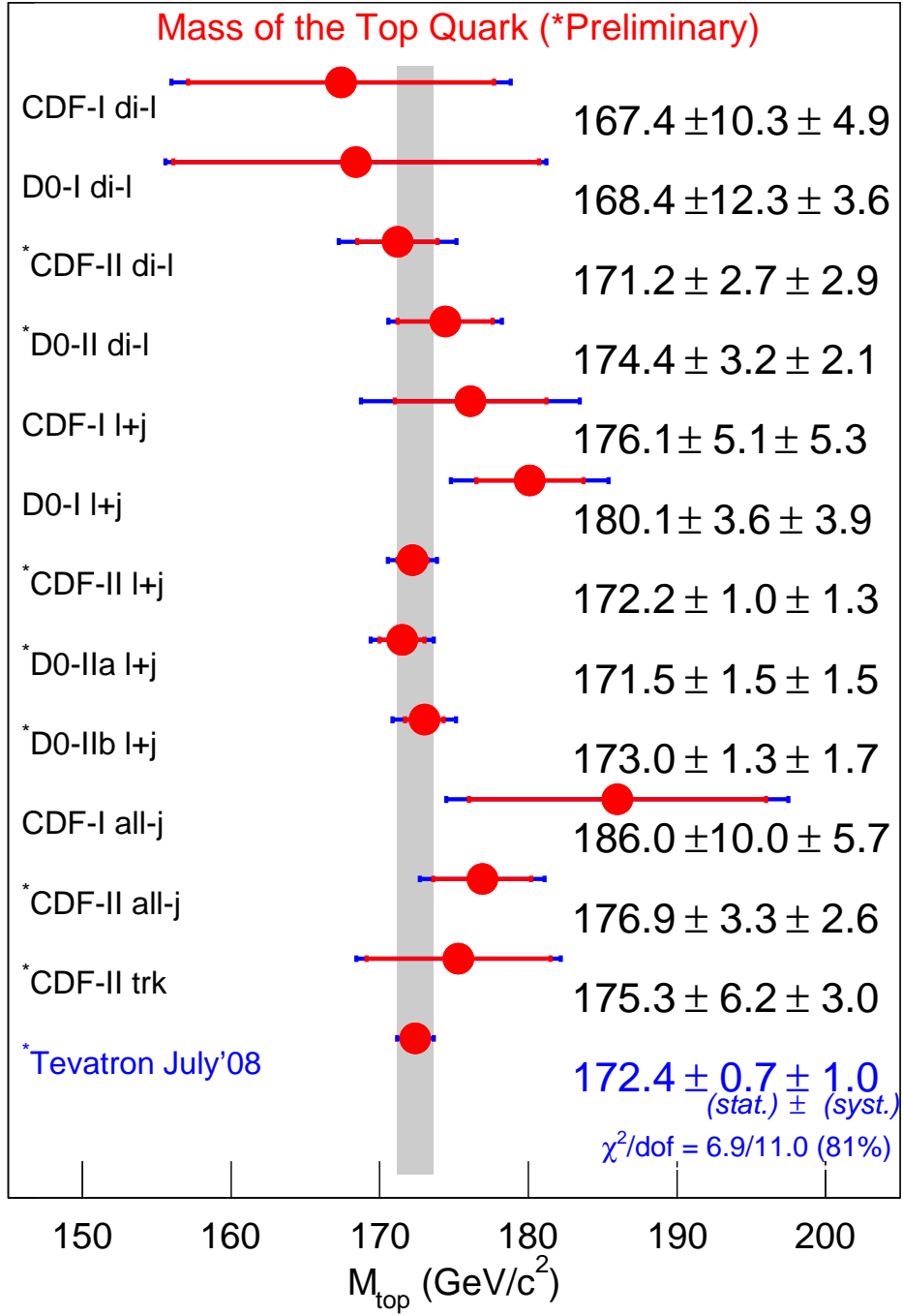}
\end{center}
\caption[Summary plot for the world average top-quark mass]
  {A summary of the input measurements and resulting world average
   mass of the top quark.}
\label{fig:summary} 
\end{figure}

The combined value for the top-quark mass is:
\begin{eqnarray}
  \MT & = & 172.4 \pm 1.2~\GeVc2\,,
\end{eqnarray}
with a $\chi^2$ of 6.9 for 11 degrees of freedom, which corresponds to
a probability of 81\%, indicating good agreement among all the input
measurements.  The total uncertainty can be sub-divided into the 
contributions from the various error categories as: Statistical ($\pm0.7$),
total JES ($\pm0.8$), Lepton scale ($\pm0.1$), Signal ($\pm0.3$), Background ($\pm0.3$), Fit
($\pm0.1$), Monte Carlo ($\pm0.3$), and UN/MI ($\pm0.02$), for a total
Systematic ($\pm1.0$), where all numbers are in units of \GeVc2.
The pull and weight for each of the inputs are listed in Table~\ref{tab:stat}.
The input measurements and the resulting world average mass of the top 
quark are summarized in Figure~\ref{fig:summary}.
\vspace*{0.10in}

The weights of many of the \RunI\ measurements are negative. 
In general, this situation can occur if the correlation between two measurements
is larger than the ratio of their total uncertainties. This is indeed the case
here.  In these instances the less precise measurement 
will usually acquire a negative weight.  While a weight of zero means that a
particular input is effectively ignored in the combination, a negative weight 
means that it affects the resulting central value and helps reduce the total
uncertainty. See reference~\cite{Lyons:1988} for further discussion of 
negative weights.

\begin{table}[t]
\begin{center}
\renewcommand{\arraystretch}{1.30}
{\small
\begin{tabular}{|l||rrr|rr||rrrr|rrr|}
\hline       
       & \multicolumn{5}{|c||}{{\RunI} published} & \multicolumn{7}{|c|}{{\RunII} preliminary} \\ \cline{2-13}
       & \multicolumn{3}{|c|}{ CDF } & \multicolumn{2}{|c||}{ D\O\ }
       & \multicolumn{4}{|c|}{ CDF } & \multicolumn{3}{|c|}{ D\O\ } \\
       & l+j     & di-l    & all-j   & l+j     & di-l    & l+j     & di-l    & all-j   & trk     & l+j/a   & l+j/b   & di-l\\
\hline
\hline
Pull   & $+0.5$  & $-0.4$  & $+1.2$  & $+1.5$  & $-0.3$  & $-0.1$  & $-0.3$  & $+1.1$  & $+0.4$    
       & $-0.5$  & $+0.3$  & $+0.5$ \\
Weight [\%]
       & $- 3.4$ & $- 0.6$ & $- 0.6$ & $+ 1.2$ & $+ 0.2$ & $+46.1$ & $+ 3.7$ & $+ 5.2$ & $+ 0.02$        
       & $+23.7$ & $+21.5$ & $+ 3.0$  \\
\hline
\end{tabular}
}
\end{center}
\caption[Pull and weight of each measurement]{The pull and weight for each of the
  inputs used to determine the world average mass of the top quark.  See 
  Reference~\cite{Lyons:1988} for a discussion of negative weights.}
\label{tab:stat} 
\end{table} 

Although the $\chi^2$ from the combination of all measurements indicates
that there is good agreement among them, and no input has an anomalously
large pull, it is still interesting to also fit for the top-quark mass
in the all-j, l+j, and di-l channels separately.  We use the same methodology,
inputs, error categories, and correlations as described above, but fit for
the three physical observables, \MTjj, \MTlj, and \MTll.
The results of this combination are shown in Table~\ref{tab:three_observables}
and have $\chi^2$ of 4.9 for 9 degrees of freedom, which corresponds to a
probability of 84\%.
These results differ from a naive combination, where
only the measurements in a given channel contribute to the \MT\ 
determination in that channel, since the combination here fully accounts
for all correlations, including those which cross-correlate the different
channels. Using the results of 
Table~\ref{tab:three_observables} we calculate the chi-squared consistency
between any two channels, including all correlations, as 
$\chi^{2}(dil-lj)=0.08$, $\chi^{2}(lj-allj)=1.7$, and 
$\chi^{2}(allj-dil)=1.9$.  These correspond to 
chi-squared probabilities of 78\%, 19\%, and 17\%, respectively, and indicate 
that the determinations of \MT\ from the three channels are consistent with 
one another.
%
%

\begin{table}[t]
\begin{center}
\renewcommand{\arraystretch}{1.30}
\begin{tabular}{|l||c|rrr|}
\hline
Parameter & Value (\GeVc2) & \multicolumn{3}{|c|}{Correlations} \\
\hline
\hline
$\MTjj$ & $177.5\pm 4.0$ & 1.00 &      &      \\
$\MTlj$ & $172.2\pm 1.2$ & 0.14 & 1.00 &      \\
$\MTll$ & $171.5\pm 2.6$ & 0.17 & 0.32 & 1.00 \\
\hline
\end{tabular}
\end{center}
\caption[Mtop in each channel]{Summary of the combination of the 12
measurements by CDF and D\O\ in terms of three physical quantities,
the mass of the top quark in the all-jets, lepton+jets, and di-lepton channels. }
\label{tab:three_observables}
\end{table}

\section{Summary}
\label{sec:summary}

A preliminary combination of measurements of the mass of the top quark
from the Tevatron experiments CDF and D\O\ is presented.  The
combination includes five published \RunI\ measurements and 
seven preliminary \RunII\ measurements.  Taking into
account the statistical and systematic uncertainties and their
correlations, the preliminary world-average result is: $\MT= 172.4 \pm
1.2~\GeVc2$, where the total uncertainty is obtained assuming Gaussian
systematic uncertainties and adding them plus the statistical
uncertainty in quadrature.  While the central value is somewhat higher
than our 2007 average, the averages are compatible as appreciably more
luminosity and refined analysis techniques are now used.
\vspace*{0.10in}

The mass of the top quark is now known with a relative precision of
0.7\%, limited by the systematic uncertainties, which are dominated by
the jet energy scale uncertainty.  This systematic is expected to
improve as larger data sets are collected since new analysis
techniques constrain the jet energy scale using in-situ $W\ra
qq^{\prime}$ decays. It can be reasonably expected that with the full
\RunII\ data set the top-quark mass will be known to better than
0.7\%.  To reach this level of precision further work is required to
determine more accurately the various correlations present, and to
understand more precisely the $b$-jet modeling, Signal, and Background
uncertainties which may limit the sensitivity at larger data sets.
Limitations of the Monte Carlo generators used to calibrate each fit
method also become more important as the precision reaches the
$\sim1~\GeVc2$ level; these warrant further study in the near future.

\clearpage

\bibliographystyle{tevewwg}
\bibliography{run2mtopJul08}

\end{document}